\documentclass[twocolumn,prl,showpacs,showkeys,amsmath,amssymb]{revtex4}
\usepackage{graphicx,amsmath,epsfig,latexsym,color}
\newcommand{\be}{\begin{equation}}
\newcommand{\ee}{\end{equation}}
\begin{document}
\title{Using Atomic Diffraction of Na from Material Gratings to Measure Atom-Surface Interactions}
\author{John D.\ Perreault, Alexander D.\ Cronin}
\affiliation{University of Arizona, Tucson, Arizona 85721}
\author{T. A. Savas}
\affiliation{Massachusetts Institute of Technology, Cambridge,
Massachusetts 02139}
\date{\today}
\begin{abstract}
In atom optics a material structure is commonly regarded as an
amplitude mask for atom waves.  However, atomic diffraction
patterns formed using material gratings indicate that material
structures also operate as phase masks.  In this study a well
collimated beam of sodium atoms is used to illuminate a silicon
nitride grating with a period of 100 nm.  During passage through
the grating slots atoms acquire a phase shift due to the van der
Waals interaction with the grating walls.  As a result the
relative intensities of the matter-wave diffraction peaks deviate
from those expected for a purely absorbing grating.  Thus a
complex transmission function is required to explain the observed
diffraction envelopes.  An optics perspective to the theory of
atomic diffraction from material gratings is put forth in the
hopes of providing a more intuitive picture concerning the
influence of the vdW potential. The van der Waals coefficient
$C_{3} = 2.7\pm 0.8\mbox{\ meV\ nm}^{3}$ is determined by fitting
a modified Fresnel optical theory to the experimental data.  This
value of $C_{3}$ is consistent with a van der Waals interaction
between atomic sodium and a silicon nitride surface.
\end{abstract}
 \maketitle

It is known that correlations of electromagnetic vacuum field
fluctuations over short distances can result in an attractive
potential between atoms.  For the case of an atom and a surface
the potential takes the form\be
\begin{split}
V(r) &= -\frac{C_{3}}{r^{3}},
\end{split}
\label{eq:vdwpot}\ee where $r$ is the atom-surface distance and
$C_{3}$ is a coefficient which describes the strength of the van
der Waals (vdW) interaction \cite{milo94}. Equation
\ref{eq:vdwpot} is often called the \emph{non-retarded} vdW
potential and is valid over distances shorter than the principle
transition wavelength of the atoms involved.  The significance of
this interaction is becoming more prevalent as mechanical
structures are being built on the nanometer scale.  The vdW
potential also plays an important part in chemistry, atomic force
microscopy, and can be used to test quantum electrodynamic theory.

Early experiments concerning the vdW interaction were based on the
deflection of atomic beams from surfaces.  It was demonstrated
that the deflection of ground state alkali \cite{shih75} and
Rydberg \cite{ande88} atom beams from a gold surface is compatible
with Eq. \ref{eq:vdwpot}.  Later measurements based on the Stark
shift interpretation of the vdW potential \cite{suke93} were
sufficiently accurate to distinguish between the retarded $V\sim
r^{-4}$ and non-retarded $V\sim r^{-3}$ forms.  More recently atom
optics techniques have been employed to measure the magnitude of
the vdW coefficient $C_{3}$.  Various ground state \cite{gris99}
and excited noble gas \cite{bruh02} atom beams have been
diffracted using nano-fabricated transmission gratings in order to
measure $C_{3}$. The influence of the vdW potential has also been
observed for large molecules in a Talbot-Lau interferometer
constructed with three gold gratings \cite{brez02}.

\begin{figure}
\scalebox{.5}{\includegraphics{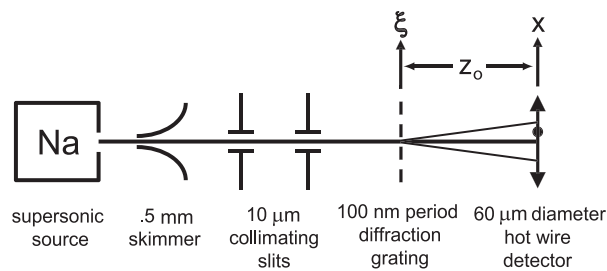}}
\caption{\label{fig:setupfig}A diagram of the experimental setup
used.}
\end{figure}

In this article we present atomic diffraction of a thermal sodium
atom beam and show that the data cannot be described by a purely
absorbing grating. A diagram of the experimental apparatus is
shown in Fig. \ref{fig:setupfig}.
The supersonic beam of sodium atoms passes through a $\mbox{.5\
mm}$ diameter skimmer and is collimated by two $\mbox{10\
}\mu\mbox{m}$ slits separated by $\sim 1\mbox{\ m}$.  By changing
the carrier gas the atom velocity can be adjusted from 0.6 to 3
km/s with $\frac{\sigma_{v}}{v}\sim .1$. The collimated atom beam
is used to illuminate a silicon nitride grating \cite{sava96} with
a period of $d = 100\mbox{\ nm}$, thickness $t = 150\pm 5\mbox{\
nm}$, open width $w = 50.5\pm 1.5\mbox{\ nm}$, and grating bar
wedge angle $\alpha = 5.25\pm .75$ degrees. All of the grating
parameters are measured independently using scanning electron
microscope images. The diffraction pattern is measured by ionizing
the sodium atoms with a hot Re wire and then counting the ions
with a channel electron multiplier.

An optical description is helpful in gaining an intuitive picture
of how the vdW interaction modifies the atomic diffraction
pattern. To this end one should recall that the Schroedinger
equation for a wave function $\psi$ can be written as\be
\begin{split}
i\hbar\frac{\partial}{\partial t}\psi(\mbox{\bf{r}},t) &=
\left[\frac{-\hbar^{2}}{2m}\nabla^{2} +
V(\mbox{\bf{r}})\right]\psi(\mbox{\bf{r}},t),
\end{split}
\label{eq:schroed}\ee where $m$ is mass, $\hbar$ is Planck's
constant, and $V$ is the potential \cite{grif95}.  One can take
the Fourier transform of Eq. \ref{eq:schroed} with respect to time
and use the fact that $\frac{\partial}{\partial
t}\Rightarrow-i\omega$ in the frequency domain to obtain\be
\begin{split}
\left[\nabla^{2} + \left(1 -
\frac{V(\mbox{\bf{r}})}{\hbar\omega}\right)k_{o}^{2}\right]\psi(\mbox{\bf{r}},\omega)
&= 0,
\end{split}
\label{eq:timeindep}\ee where the dispersion relation
$\omega=\frac{\hbar k_{o}^{2}}{2m}$ has been utilized.  Equation
\ref{eq:timeindep} is usually referred to as the time independent
Schroedinger equation.  It is quite illuminating to recall that
the Helmholtz equation \cite{jack99} for the electric field
$\mbox{\bf{E}}$ is given by\be
\begin{split}
\left(\nabla^{2} +
n^{2}k_{o}^{2}\right)\mbox{\bf{E}}(\mbox{\bf{r}},\omega) = 0,
\end{split}
\label{eq:helm}\ee where $n$ is index of refraction. By inspection
one can see that Eqs. \ref{eq:timeindep} and \ref{eq:helm} are
formally equivalent where the quantities $n$ and $\left(1 -
\frac{V}{\hbar\omega}\right)^{1/2}$ play analogous roles. Due to
this fact many wave propagation methods developed in optics can be
applied directly to matter wave propagation, being mindful of the
fact that in optics $\omega=ck_{o}$.

While Eq. \ref{eq:timeindep} can be formally solved using a
Green's function approach, approximate solutions used in physical
optics can lead to a better understanding of how the vdW
interaction affects atomic diffraction patterns. The Fresnel and
Fraunhofer approximations are commonly used in optics and
represent a useful tool when faced with propagating the wave
function $\psi$ from the grating to the detector plane.  The
Fresnel or paraxial approximation is valid as long as the
propagation distance $z$ satisfies the inequality\be
\begin{split}
z\gg|x-\xi|.
\end{split}
\label{eq:fresnel}\ee This is certainly satisfied for our
experiment since the diffraction angles are less than $10^{-3}$
radians and the orders are resolved.  The Fraunhofer or far-field
approximation goes beyond the Fresnel approximation by requiring
that\be
\begin{split}
z\gg\frac{k_{o}}{2}\xi_{max}^{2} =
\frac{\pi}{\lambda_{dB}}\xi_{max}^{2},
\end{split}
\label{eq:fraun}\ee where $\lambda_{dB}$ is the de Brolglie
wavelength of the atoms and $\xi_{max}$ is the relevant extent in
the aperture plane \cite{good96}.  For the case of propagation
from a uniformly illuminated grating of period $d$ to the detector
plane, $\xi_{max}\rightarrow d$ and Eq. \ref{eq:fraun} takes the
form $z\gg\frac{\pi d^{2}}{\lambda_{dB}}$. For our experimental
setup $d=100\mbox{\ nm}$ and $\lambda_{dB}\sim 10^{-11}\mbox{\
m}$, so the inequality $z\approx 2\mbox{\
m}\gg\frac{\pi}{1000}\mbox{\ m}$ is met.  However, our atom beam
diameter is on the order of $10^{-5}$ m and so
$\xi_{max}\rightarrow 10^{-5}$ m implying that the inequality in
Eq. \ref{eq:fraun} is not met.

In light of the previous discussion it seems most appropriate to
use the Fresnel approximation to model our experiment.  According
to the Fresnel approximation the wave function in the detector
plane $\psi(x)$ is related to that just after the grating
$\psi(\xi)$ by a scaled spatial Fourier transform\be
\begin{split}
\psi(x) &\propto
\left.\mathcal{F}\left\{e^{i\frac{k_{o}\xi^{2}}{2z_{o}}}\psi(\xi)\right\}\right|_{f_{\xi}=\frac{x}{\lambda_{dB}z_{o}}},
\end{split}
\label{eq:fresprop}\ee where $\mathcal{F}\{\}$ denotes a Fourier
transform and $f_{\xi}$ is the Fourier conjugate variable to $\xi$
\cite{good96}.  The quadratic phase factor in Eq.
\ref{eq:fresprop} accounts for the fact that the phase fronts have
a parabolic shape before the far-field is reached.

The wave function just after the grating $\psi(\xi)$ is given
by\be
\begin{split}
\psi(\xi) &= \left[T(\xi)
\ast\mbox{comb}\left(\frac{\xi}{d}\right)\right] U(\xi),
\end{split}
\label{eq:trans}\ee where $\mbox{comb}\left(\frac{\xi}{d}\right)$
is an array of delta functions with spacing $d$, the operator
$\ast$ denotes a convolution, and $U(\xi)$ is complex function
describing the atom beam amplitude in the plane of the grating.
The transmission function of a single grating window $T(\xi)$ in
Eq. \ref{eq:trans} is defined as \be
\begin{split}
T(\xi)&\equiv e^{i\phi(\xi)}\mbox{rect}\left(\frac{\xi}{w}\right),
\end{split}
\label{eq:sstrans}\ee where $\mbox{rect}(arg)=1$ when
$|arg|\leq\frac{1}{2}$ and zero otherwise.  The phase $\phi(\xi)$
accounts for the vdW interaction and its origin will be discussed
later.  This description of $\psi(\xi)$ and $T(\xi)$ in terms of
the functions comb() and rect() is standard Fourier optics
notation and convenient due to its modular nature \cite{good96}.

Equation \ref{eq:trans} can then be substituted into Eq.
\ref{eq:fresprop} to obtain\be
\begin{split}
\psi(x) &\propto \sum_{j=-\infty}^{\infty}
\mathcal{A}_{j}U\left(x-j\frac{\lambda_{dB}z_{o}}{d}\right),
\end{split}
\label{eq:psidet}\ee where the summation index corresponds to the
$j^{th}$ diffraction order, the diffraction amplitude
$\mathcal{A}_{j}$ is defined as\be
\begin{split}
\mathcal{A}_{j} &\equiv
\left.\mathcal{F}\left\{T(\xi)\right\}\right|_{f_{\xi}=\frac{j}{d}}
=\left.\mathcal{F}\left\{e^{i\phi(\xi)}\mbox{rect}\left(\frac{\xi}{w}\right)\right\}\right|_{f_{\xi}=\frac{j}{d}},
\end{split}
\label{eq:diffenv}\ee and the beam profile in the detector plane
is given by\be
\begin{split}
U(x) =
\left.\mathcal{F}\left\{e^{i\frac{k_{o}\xi^{2}}{2z_{o}}}U(\xi)\right\}\right|_{f_{\xi}=\frac{x}{\lambda_{dB}z_{o}}}.
\end{split}
\label{eq:beamdet}\ee From Eq. \ref{eq:psidet} we can predict the
atom intensity\be
\begin{split}
I(x)\equiv\left|\psi(x)\right|^{2},
\end{split}
\label{eq:detint}\ee in the detector plane which can also be
interpreted as the probability law for atoms.  A distribution of
atom velocities can be incorporated by a weighted incoherent sum
of the intensity pattern for each atom velocity
$I\left(x;v\right)$\be
\begin{split}
I(x) &= \sum_{v}P(v)I\left(x;v\right);\qquad v =
\frac{h}{m\lambda_{dB}},
\end{split}
\label{eq:dispint}\ee\be
\begin{split}
P(v) &\propto v^{3}exp\left(-\frac{m(v-u)^{2}}{2k_{B}T}\right),
\end{split}
\label{eq:pv}\ee where the $P(v)$ is the probability distribution
function of velocities for a supersonic source, $u$ is the average
flow velocity, $k_{B}$ is Boltzmann's constant, and $T$ is the
longitudinal temperature of the beam in the moving frame of the
atoms \cite{dunn96}.

One can see from Eq. \ref{eq:psidet} that the diffraction pattern
consists of replications of the beam shape $\left|U(x)\right|^{2}$
shifted by integer multiples of $\frac{\lambda_{dB}z_{o}}{d}$ with
relative intensities determined by the modulus squared of Eq.
\ref{eq:diffenv}.  An important feature to notice in Eq.
\ref{eq:diffenv} is that a diffraction order in the detector plane
corresponds to a spatial frequency in the grating plane through
the relation $f_{\xi}=\frac{j}{d}$. This highlights the connection
between the spatially dependent phase $\phi(\xi)$ in Eq.
\ref{eq:sstrans} and the magnitude of the diffraction orders in
Eq. \ref{eq:psidet}.

The earlier assertion that $\phi(\xi)$ in Eq. \ref{eq:sstrans}
somehow incorporates the vdW interaction into the optical
propagation theory can be understood by recalling from Eq.
\ref{eq:timeindep} that the index of refraction $n$ and quantity
$\left(1 - \frac{V}{\hbar\omega}\right)^{1/2}$ play similar roles
in optics and atom optics, respectively. In optics one calculates
a phase shift $\phi$ induced by a glass plate by multiplying the
wavenumber in the material $nk_{o}$ by the thickness of the plate
$L$ (i.e. $\phi=nk_{o}L$).  Just as in the optics case one can
calculate the phase shift $\phi(\xi)$ accumulated by the wave
function passing through the grating windows\be
\begin{split}
\phi(\xi) &= \int\left(\parbox{.75in}{wavenumber in
potential}\right) \left(\parbox{.65in}{differential
thickness}\right)\\ &=
\int_{-t}^{0}k_{o}\left(1-\frac{V(\xi,z)}{\hbar\omega}\right)^{1/2}dz,
\end{split}
\label{eq:fullwkb}\ee where $t$ is the thickness of the grating
and $V(\xi,z)$ is the potential the atoms experience between the
grating bars due to the vdW interaction. Thus the vdW interaction
is analogous to a glass plate with a spatially dependent index of
refraction, a kind of diverging lens that fills each grating
window. The result in Eq. \ref{eq:fullwkb} is consistent with the
wave function phase according to the WKB approximation
\cite{grif95}.

In arriving at Eq. \ref{eq:fullwkb} diffraction due to abrupt
changes in the potential $V(\xi,z)$ has been ignored while the
wave function propagates through the grating windows. This is a
valid approximation due to the fact that $\lambda_{dB}\ll
w,\left[\frac{\partial}{\partial\xi}\frac{V(\xi)}{\hbar\omega}\right]^{-1}$
in the region of the potential that corresponds to the diffraction
orders of interest.  The relationship between spatial regions of
the potential $V(\xi,z)$ and a given diffraction order will be
discussed in subsequent paragraphs.  It is also important to note
that Eq. \ref{eq:fullwkb} assumes that the potential $V(\xi,z)$
exists only between the grating bars (i.e. $V(\xi,z)=0$ for $z<-t$
or $z>0$) and neglects the fact that the bars are not
semi-infinite planes.  Theoretical work done by Spurch et al.
\cite{ZhS95} suggests that the vdW potential corresponding to our
$\sim 50$ nm grating bar width is very similar to that of a
semi-infinite plane in the $\xi$ direction.  Since the phase
$\phi(\xi)$ from Eq. \ref{eq:fullwkb} only depends on the integral
of the potential in the $z$ direction one would also expect that
edge effects in $V(\xi,z)$ due to the finite grating thickness $t$
are a small correction.

If the particle energy $\hbar\omega$ is much greater than the
potential $V(\xi,z)$ then Eq. \ref{eq:fullwkb} can be further
simplified by Taylor expanding the quantity $\left(1 -
\frac{V}{\hbar\omega}\right)^{1/2}$ and keeping the leading order
term in $\frac{V}{\hbar\omega}$\be
\begin{split}
\phi(\xi) = tk_{o}-\frac{1}{\hbar v}\int_{-t}^{0}V(\xi,z)dz;\qquad
\hbar\omega\gg V(\xi,z),
\end{split}
\label{eq:firstwkb}\ee through the use of the dispersion relation
$\omega=\frac{\hbar k_{o}^{2}}{2m}$ and $p=mv=\hbar k_{o}$.
Equation \ref{eq:firstwkb} is often called the Eikonal
approximation.  The term $tk_{o}$ in Eq. \ref{eq:firstwkb} is
independent of $\xi$ and of no consequence in Eq. \ref{eq:diffenv}
so it can be neglected. One can see from Eq. \ref{eq:firstwkb}
that if $V(\xi,z)\rightarrow 0$ then Eq. \ref{eq:diffenv} reduces
to the sinc diffraction envelope expected from a purely absorbing
grating. Furthermore, it is now clear from Eqs. \ref{eq:diffenv}
and \ref{eq:firstwkb} that the relative heights of the diffraction
orders are altered in a way that depends on $V(\xi,z)$ as well as
the atom beam velocity $v$.

As a simple model one can represent the potenial in Eq.
\ref{eq:firstwkb} as the sum of the potential due to the two
interior walls of the grating window\be
\begin{split}
\phi(\xi) = -\frac{t}{\hbar
v}\left[g_{-}(\alpha)V_{-}(\xi)-g_{+}(\alpha)V_{+}(\xi)\right],
\end{split}
\label{eq:winphase}\ee where the function $g_{\pm}(\alpha)$
incorporates the influence of the wedge angle $\alpha$\be
\begin{split}
g_{\pm}(\alpha)\equiv\frac{1\pm\frac{t\tan\alpha}{2\left(\xi\pm\frac{w}{2}\right)}}{\left(1\pm\frac{t\tan\alpha}{\left(\xi\pm\frac{w}{2}\right)}\right)^{2}},
\end{split}
\label{eq:galpha}\ee and $V_{\pm}(\xi)\equiv
C_{3}\left(\xi\pm\frac{w}{2}\right)^{-3}$ is implied by Eq.
\ref{eq:vdwpot}.  Equations \ref{eq:winphase} and \ref{eq:galpha}
are arrived at by carrying out the integration in Eq.
\ref{eq:firstwkb} while assuming that the open grating width $w$
varies in the propagation direction $z$ as $w(z)=w+2z\tan\alpha$.
Since the principle transition wavelength of Na (590 nm) is much
larger than $\frac{w}{2}$ (i.e. the maximum atom-surface distance
of $\sim 25$ nm) the non-retarded form of the vdW potential is
appropriate.

\begin{figure}
\scalebox{.65}{\includegraphics{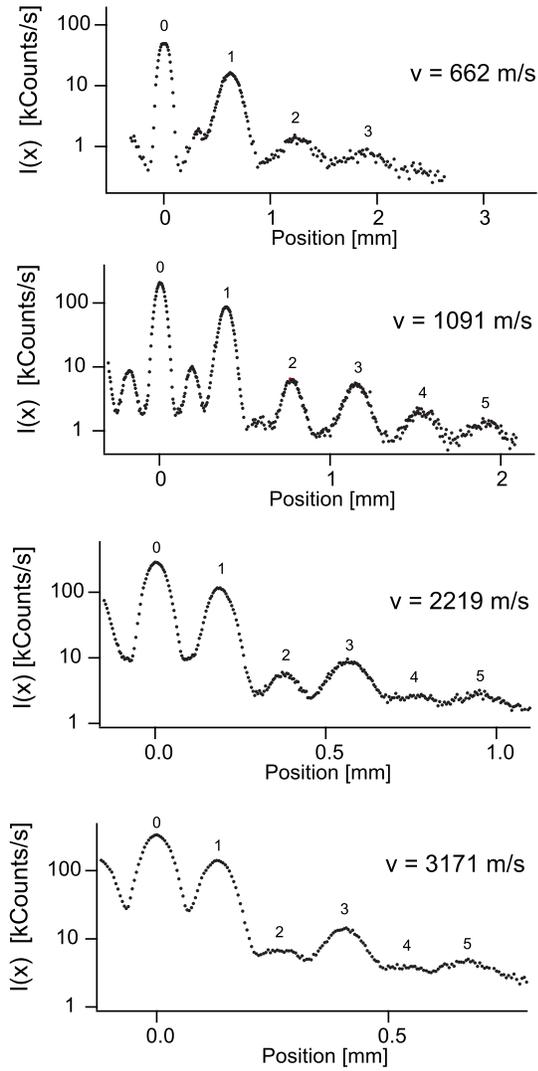}}
\caption{\label{fig:diffig}Observed diffraction patterns of four
different atom velocities.  The numbers next to the peaks indicate
the diffraction order.  Molecular $\mbox{Na}_{2}$ peaks are also
visible between zeroth and first orders for slower velocities.}
\end{figure}

It is not immediately obvious how the phase representation in Eq.
\ref{eq:firstwkb} will affect the far-field diffraction pattern or
if the Eikonal approximation is appropriate in light of Eq.
\ref{eq:winphase} (i.e. $V_{\pm}(\xi)\rightarrow -\infty$ as $|\xi
|\rightarrow\frac{w}{2}$). In order to address this it is helpful
to introduce the concept of an \emph{instantaneous} spatial
frequency \cite{boyd92}\be
\begin{split}
f_{\xi}\left(\xi_{j}\right)
\equiv\left.\frac{\partial\phi}{\partial\xi}\right|_{\xi=\xi_{j}}
= \frac{j}{d},
\end{split}
\label{eq:instfreq}\ee where $\xi_{j}$ is the grating window
location corresponding to the diffraction order $j$ as in Eq.
\ref{eq:diffenv}. For the limiting case of $\alpha\rightarrow 0$
the geometry factor $g_{\pm}(\alpha)\rightarrow 1$ the higher
order terms in Eq. \ref{eq:fullwkb} will become important when
$\xi_{j}\rightarrow\xi_{c}$ and
$C_{3}\left(\xi_{c}-\frac{w}{2}\right)^{-3}\approx\hbar\omega$. If
Eq. \ref{eq:winphase} is inserted into Eq. \ref{eq:instfreq} with
the previously mentioned limits one can solve for the diffraction
order $j_{c}$ at which the approximation in Eq. \ref{eq:firstwkb}
breaks down\be
\begin{split}
j_{c} &\approx
\frac{3k_{o}t}{2}\frac{C_{3}d}{\hbar\omega\left(\xi_{c}-\frac{w}{2}\right)^{4}}
=
\frac{3k_{o}t}{2}\left(\frac{d^{3}\hbar\omega}{C_{3}}\right)^{\frac{1}{3}}.
\end{split}
\label{eq:critorder}\ee For the present experiment
$\frac{3k_{o}t}{2}\sim 10^{5}$ and
$\left(\frac{d^{3}\hbar\omega}{C_{3}}\right)^{\frac{1}{3}}\sim
10^{\frac{7}{3}}$ which implies that $j_{c}\sim 10^{7}$. Thus the
approximation in Eq. \ref{eq:firstwkb} is appropriate since we
typically concerned with only the first ten diffraction orders. In
fact, the paraxial approximation will become invalid before Eq.
\ref{eq:firstwkb} becomes invalid due to the fact the diffraction
order spacing is typically $\frac{\lambda_{dB}z_{o}}{d}\sim
.1\mbox{\ mm}$.  It is also interesting to note that using Eqs.
\ref{eq:winphase} and \ref{eq:instfreq} one can solve for the
position $\xi_{j}$ in the grating window\be
\begin{split}
\xi_{j}\approx\frac{w}{2} - \left(\frac{3tC_{3}d}{j\hbar
v}\right)^{\frac{1}{4}};\quad j\geq 1,
\end{split}
\label{eq:xifreq}\ee corresponding to a particular diffraction
order $j$.  If $j=1$ in Eq. \ref{eq:xifreq} then $\xi_{1}\approx
3.2$ nm and since $\xi_{j}\sim j^{-\frac{1}{4}}$ the shape of the
diffraction amplitude in Eq. \ref{eq:diffenv} depends on a small
region of the potential near an an atom-surface distance of $\sim
20$ nm.

The experimental data for diffraction patterns of four different
atom beam velocities are displayed in Fig. \ref{fig:diffig}.  One
can see from Fig. \ref{fig:diffig} that the second order
diffraction peak is almost completely suppressed for the faster
atoms whereas it is quite pronounced for the slower atoms.  This
velocity dependence is a clear indication that a complex
transmission functions such as Eq. \ref{eq:sstrans} (i.e. $C_{3}
\neq 0$) is required to explain the data. A least-squares fit to
Eqs. \ref{eq:psidet} and \ref{eq:dispint} is used to determine
diffraction envelope $\left|\mathcal{A}_{j}\right|^{2}$ and the
average velocity. It is clear from Fig. \ref{fig:diffig} that the
diffraction orders overlap to some extent, hence the tails of the
beam shape are important when determining
$\left|\mathcal{A}_{j}\right|^{2}$. The broad tails of the beam
shape were not adequately described by a Gaussian so an empirical
shape using a fixed collimating geometry was derived from the
measured raw beam profile and used for $\left|U(x)\right|^{2}$.

\begin{figure}
\scalebox{.66}{\includegraphics{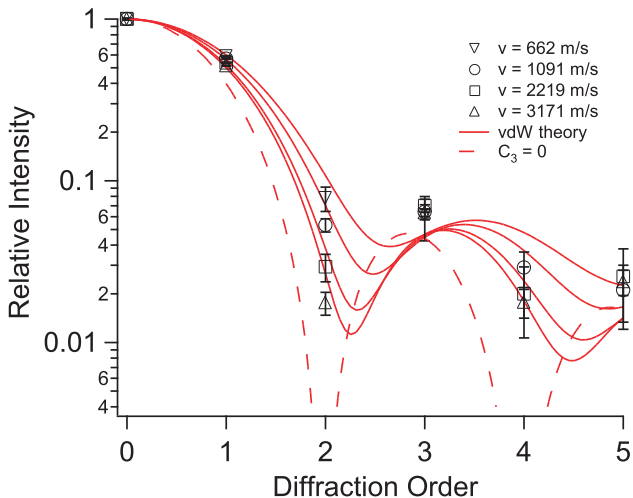}}
\caption{\label{fig:envfig}Diffraction order intensities and best
fit diffraction amplitude $\left|\mathcal{A}_{j}\right|^{2}$.
Notice how the theory for a purely absorbing grating fails to
describe the data.}
\end{figure}

The diffraction amplitudes $\left|\mathcal{A}_{j}\right|^{2}$
determined from Fig. \ref{fig:diffig} for the various velocities
are displayed in Fig. \ref{fig:envfig}. The vdW coefficient
$C_{3}$ is determined by a least-squares fit to this reduced data
with the modulus squared of Eq. \ref{eq:diffenv}.  All of the
grating parameters are determined independently, therefore $C_{3}$
is the only free parameter.  Data from each velocity is fit
simultaneously with the same $C_{3}$. It is clear that a purely
absorbing grating (i.e. $C_{3}=0$) is inconsistent with all of the
observed $\left|\mathcal{A}_{j}\right|^{2}$ especially at lower
velocities for which the phase $\phi(\xi)$ is much larger.
Uncertainty in the determination of the grating parameter $w$ and
the exact shape of the potential in Eq. \ref{eq:firstwkb} may be
responsible for the slight deviation from theory evident in Fig.
\ref{fig:envfig}.

A study of the systematic errors in our experiment and analysis
suggest that $w$ is largest source of uncertainty when calculating
$C_{3}$.  One can numerically calculate the function ${C}_{3}(w)$,
which is the best fit $C_{3}$ as a function of $w$, whose linear
dependence around the physical value of $w$ is found to be
$\left.\frac{\partial C_{3}}{\partial w}\right|_{w=50.5\mbox{\
}nm} = .52\mbox{\ }\mbox{\ meV\ nm}^{2}$. The error in $C_{3}$ is
arrived at by taking the product of this slope and the 1.5 nm
uncertainty in $w$. After carrying out the previously described
analysis we obtain a value for the vdW coefficient $C_{3} = 2.7\pm
0.8\mbox{\ }\mbox{\ meV\ nm}^{3}$.  The uncertainty determined
this way is considerably larger than the statistical uncertainty
in $C_{3}$ from the least-squares fitting procedure.  The
uncertainty due to $w$ is also larger than the systematic
corrections due to the atom beam profile or uncertainties due to
imperfect knowledge of the grating parameters: $d$, $t$, and
$\alpha$.

\begin{table}[h!]
\caption{\label{tab:C3} Measured and calculated values of $C_3$}
\begin{ruledtabular}
\begin{tabular}{ll}
  Method & $C_3 \ \left[\mbox{\ meV\ nm}^{3}\right]$\\  \hline
  This experiment & $2.7 \pm 0.8$ \\
  Na and perfect conductor \cite{DJS99}   & 7.60  \\
  Na$^\dagger$ and perfect conductor \cite{MDB97,Lif56,MeS98} & 6.29  \\
  Na$^\dagger$ and  Na surface     & 4.1  \\
  Na$^\dagger$ and  SiN$_x$ surface   & 3.2  \\
  Na$^\dagger$ and  SiN$_x$  with a 1-nm Na layer$^\S$ & 3.8  \\
\end{tabular}
\end{ruledtabular}
\noindent $^\dagger$ indicates a one-oscillator model for atomic
polarizability. \\ $^\S$ indicates $C_3$ evaluated $20$ nm from
the first surface.
\end{table}

To compare our experimental measurement with theoretical
predictions of the van der Waal potential strength, we evaluate
five different theoretical cases for sodium atoms and various
surfaces in Table \ref{tab:C3}. The Lifshitz formula \cite{Lif56}
for $C_3$ is
\begin{equation} C_3 = \frac{\hbar}{4\pi}\int_0^{\infty} d \omega
\alpha(i \omega) \frac{\epsilon(i\omega)-1} {\epsilon(i\omega)+1},
\label{eq.lifshitz} \end{equation} where $\alpha(i \omega)$ is the
dynamic polarizability of the atom and $\epsilon(iw)$ is the
permittivity of the surface material, both of which are a function
of complex frequency.

A single Lorentz oscillator model for an atom (i.e. neglecting all
but the valence electron) with no damping gives an expression for
polarizability \cite{MeS98}
\begin{equation} \alpha(i\omega) =  \frac {\alpha(0)}
{1+ (\frac{\omega}{\omega_0})^2}.\label{eq:pol}\end{equation} For
sodium atoms $\alpha(0)=24.1 \ \AA^3$ \cite{ESC95} and $\omega_0 =
2\pi c / (590 \ \mbox{nm})$.  Combining this with a perfect
conductor (i.e. $\epsilon = \infty$) in Eq. \ref{eq.lifshitz}
gives $C_3= 6.29 \ \mbox{meV nm}^3$. This value agrees well with
the non-retarded limit calculated in reference \cite{MDB97} for
sodium atoms with a single valence electron.

For more accurately modeled sodium atoms and a perfect conductor,
Derevianko $et$ $al.$ \cite{DJS99} calculated $C_3= 7.60 \
\mbox{meV nm}^3$ and reported a range of values spanning $0.08 \
\mbox{meV nm}^3$ based on different many-body calculation methods
which all include the effect of core electrons. It is noteworthy
that $16\%$ of this recommended value is due to the core electrons
\cite{DJS99}.

For a metal surface, the Drude model describes $\epsilon(i\omega)$
in terms of the plasma frequency and damping:
\begin{equation} \epsilon(i \omega) = 1 + \frac{\omega_p^2}
{\omega(\omega + \gamma)}. \end{equation} For sodium metal,
$\hbar \omega_p = 5.8 \ \mbox{eV}$ and $\hbar \gamma = 23 \ \mbox{meV},$ 
resulting in $C_3 = 4.1 \ \mbox{meV nm}^3$ for a sodium atom and a
bulk sodium surface.  Presumably this calculation also
under-estimates $C_3$ because the core electrons are neglected.
However, the calculation error is probably smaller than that of a
perfect conductor because the core electron excitations are at
frequencies comparable to $\omega_p$.

For an insulating surface of silicon nitride, which is the
diffraction grating material, Bruhl $et$ $al$ \cite{bruh02} used a
model with
\begin{equation} \epsilon(i \omega) = \frac{\omega^2 +
(1+g_0)\omega_0^2} {\omega^2 + (1-g_0)\omega_0^2}
\label{eq:bruhl}\end{equation} where $\hbar \omega_0 \equiv E_s =
13 \mbox{eV}$ and $g_0 = 0.588$ is the material response function
at zero frequency. Using Eqs. \ref{eq.lifshitz}, \ref{eq:pol}, and
\ref{eq:bruhl} gives a value of $C_3 = 3.2 \ \mbox{meV nm}^3$.

A multilayered surface makes a vdW potential that no longer
depends exactly on $r^{-3},$ even in the non-retarded limit. We
used Equation 4.10 from reference \cite{ZhS95} to calculate $V(r)$
for thin films of sodium on a slab of silicon nitride. Because our
experiment is sensitive to atom-surface distances in the region 20
nm, we report the nominal value of $C_3$ from these calculations
using $C_3 = V(\mbox{20 nm})\times (\mbox{20 nm})^3$.  Evaluated
this way, isolated thin films make a smaller $C_3$ as $r$
increases. Films on a substrate make $C_3$ vary from the value
associated with the bulk film material to the value associated
with the bulk substrate material as $r$ increases.

In conclusion an optics perspective to the theory of atomic
diffraction from a material grating has been put forth. The
results in Eqs. \ref{eq:diffenv}, \ref{eq:firstwkb} and
\ref{eq:winphase} have been derived using Fourier optics
techniques and appear to be consistent with the diffraction theory
presented in \cite{gris00}. Diffraction data for a sodium atom
beam at four different velocities show clear evidence of
atom-surface interactions with the silicon nitride grating.  A
complex transmission function such as that in Eq. \ref{eq:sstrans}
is required to explain the data.  The measured value of $C_{3} =
2.7\pm 0.8\mbox{\ }\mbox{\ meV\ nm}^{3}$ is limited in precision
by uncertainty of the grating parameter $w$. Based on the results
in Table \ref{tab:C3} for a single Lorentz oscillator the new
measurement of $C_{3}$ presented in this article is consistent
with a vdW interaction between atomic sodium and a silicon nitride
surface. Our measurement is inconsistent with a perfectly
conducting surface and also a silicon nitride surface coated with
more than one nm of bulk sodium. This implies that atomic
diffraction from a material grating may provide the means to test
the theory of vdW interactions with a multi-layered surface
\cite{ZhS95} by using coated gratings.

The authors would like to thank Hermann Uys for technical
assistance.


\begin{thebibliography}{20}
\expandafter\ifx\csname
natexlab\endcsname\relax\def\natexlab#1{#1}\fi
\expandafter\ifx\csname bibnamefont\endcsname\relax
  \def\bibnamefont#1{#1}\fi
\expandafter\ifx\csname bibfnamefont\endcsname\relax
  \def\bibfnamefont#1{#1}\fi
\expandafter\ifx\csname citenamefont\endcsname\relax
  \def\citenamefont#1{#1}\fi
\expandafter\ifx\csname url\endcsname\relax
  \def\url#1{\texttt{#1}}\fi
\expandafter\ifx\csname
urlprefix\endcsname\relax\def\urlprefix{URL }\fi
\providecommand{\bibinfo}[2]{#2}
\providecommand{\eprint}[2][]{\url{#2}}

\bibitem[{\citenamefont{Milonni}(1994)}]{milo94}
\bibinfo{author}{\bibfnamefont{P.~W.} \bibnamefont{Milonni}},
  \emph{\bibinfo{title}{The Quantum Vacuum}} (\bibinfo{publisher}{Academic
  Press}, \bibinfo{year}{1994}).

\bibitem[{\citenamefont{Shih and Parsegian}(1975)}]{shih75}
\bibinfo{author}{\bibfnamefont{A.}~\bibnamefont{Shih}} \bibnamefont{and}
  \bibinfo{author}{\bibfnamefont{V.~A.} \bibnamefont{Parsegian}},
  \bibinfo{journal}{Phys. Rev. A} \textbf{\bibinfo{volume}{12}},
  \bibinfo{pages}{835} (\bibinfo{year}{1975}).

\bibitem[{\citenamefont{Anderson et~al.}(1988)\citenamefont{Anderson, Haroche,
  Hinds, W., and Meschede}}]{ande88}
\bibinfo{author}{\bibfnamefont{A.}~\bibnamefont{Anderson}},
  \bibinfo{author}{\bibfnamefont{S.}~\bibnamefont{Haroche}},
  \bibinfo{author}{\bibfnamefont{E.~A.} \bibnamefont{Hinds}},
  \bibinfo{author}{\bibfnamefont{J.}~\bibnamefont{W.}}, \bibnamefont{and}
  \bibinfo{author}{\bibfnamefont{D.}~\bibnamefont{Meschede}},
  \bibinfo{journal}{Phys. Rev. A} \textbf{\bibinfo{volume}{37}},
  \bibinfo{pages}{3594} (\bibinfo{year}{1988}).

\bibitem[{\citenamefont{Sukenik et~al.}(1993)\citenamefont{Sukenik, Boshier,
  Cho, Sandoghdar, and Hinds}}]{suke93}
\bibinfo{author}{\bibfnamefont{C.~I.} \bibnamefont{Sukenik}},
  \bibinfo{author}{\bibfnamefont{M.~G.} \bibnamefont{Boshier}},
  \bibinfo{author}{\bibfnamefont{D.}~\bibnamefont{Cho}},
  \bibinfo{author}{\bibfnamefont{V.}~\bibnamefont{Sandoghdar}},
  \bibnamefont{and} \bibinfo{author}{\bibfnamefont{E.~A.} \bibnamefont{Hinds}},
  \bibinfo{journal}{Phys. Rev. Lett.} \textbf{\bibinfo{volume}{70}},
  \bibinfo{pages}{560} (\bibinfo{year}{1993}).

\bibitem[{\citenamefont{Grisenti et~al.}(1999)\citenamefont{Grisenti,
  Schollkopf, Toennies, Hegerfeldt, and Kohler}}]{gris99}
\bibinfo{author}{\bibfnamefont{R.~E.} \bibnamefont{Grisenti}},
  \bibinfo{author}{\bibfnamefont{W.}~\bibnamefont{Schollkopf}},
  \bibinfo{author}{\bibfnamefont{J.~P.} \bibnamefont{Toennies}},
  \bibinfo{author}{\bibfnamefont{G.~C.} \bibnamefont{Hegerfeldt}},
  \bibnamefont{and} \bibinfo{author}{\bibfnamefont{T.}~\bibnamefont{Kohler}},
  \bibinfo{journal}{Phys. Rev. Lett.} \textbf{\bibinfo{volume}{83}},
  \bibinfo{pages}{1755} (\bibinfo{year}{1999}).

\bibitem[{\citenamefont{Bruhl et~al.}(2002)\citenamefont{Bruhl, Fouquet,
  Grisenti, Toennies, Hegerfeldt, Kohler, Stoll, and Walter}}]{bruh02}
\bibinfo{author}{\bibfnamefont{R.}~\bibnamefont{Bruhl}},
  \bibinfo{author}{\bibfnamefont{P.}~\bibnamefont{Fouquet}},
  \bibinfo{author}{\bibfnamefont{R.~E.} \bibnamefont{Grisenti}},
  \bibinfo{author}{\bibfnamefont{J.~P.} \bibnamefont{Toennies}},
  \bibinfo{author}{\bibfnamefont{G.~C.} \bibnamefont{Hegerfeldt}},
  \bibinfo{author}{\bibfnamefont{T.}~\bibnamefont{Kohler}},
  \bibinfo{author}{\bibfnamefont{M.}~\bibnamefont{Stoll}}, \bibnamefont{and}
  \bibinfo{author}{\bibfnamefont{D.}~\bibnamefont{Walter}},
  \bibinfo{journal}{Europhys. Lett.} \textbf{\bibinfo{volume}{59}},
  \bibinfo{pages}{357} (\bibinfo{year}{2002}).

\bibitem[{\citenamefont{Brezger et~al.}(2002)\citenamefont{Brezger,
  Hackermuller, Uttenthaler, Petschinka, Arndt, and Zeilinger}}]{brez02}
\bibinfo{author}{\bibfnamefont{B.}~\bibnamefont{Brezger}},
  \bibinfo{author}{\bibfnamefont{L.}~\bibnamefont{Hackermuller}},
  \bibinfo{author}{\bibfnamefont{S.}~\bibnamefont{Uttenthaler}},
  \bibinfo{author}{\bibfnamefont{J.}~\bibnamefont{Petschinka}},
  \bibinfo{author}{\bibfnamefont{M.}~\bibnamefont{Arndt}}, \bibnamefont{and}
  \bibinfo{author}{\bibfnamefont{A.}~\bibnamefont{Zeilinger}},
  \bibinfo{journal}{Phys. Rev. Lett.} \textbf{\bibinfo{volume}{88}},
  \bibinfo{pages}{100404} (\bibinfo{year}{2002}).

\bibitem[{\citenamefont{Savas et~al.}(1996)\citenamefont{Savas, Schattenburg,
  Carter, and Smith}}]{sava96}
\bibinfo{author}{\bibfnamefont{T.~A.} \bibnamefont{Savas}},
  \bibinfo{author}{\bibfnamefont{M.~L.} \bibnamefont{Schattenburg}},
  \bibinfo{author}{\bibfnamefont{J.~M.} \bibnamefont{Carter}},
  \bibnamefont{and} \bibinfo{author}{\bibfnamefont{H.~I.} \bibnamefont{Smith}},
  \bibinfo{journal}{J. Vac. Sci. Tech. B} \textbf{\bibinfo{volume}{14}},
  \bibinfo{pages}{4167} (\bibinfo{year}{1996}).

\bibitem[{\citenamefont{Griffiths}(1995)}]{grif95}
\bibinfo{author}{\bibfnamefont{D.~J.} \bibnamefont{Griffiths}},
  \emph{\bibinfo{title}{Introduction to Quantum Mechanics}}
  (\bibinfo{publisher}{Prentice Hall}, \bibinfo{year}{1995}).

\bibitem[{\citenamefont{Jackson}(1999)}]{jack99}
\bibinfo{author}{\bibfnamefont{J.~D.} \bibnamefont{Jackson}},
  \emph{\bibinfo{title}{Classical Electrodynamics}} (\bibinfo{publisher}{John
  Wiley \& Sons}, \bibinfo{year}{1999}).

\bibitem[{\citenamefont{Goodman}(1996)}]{good96}
\bibinfo{author}{\bibfnamefont{J.~W.} \bibnamefont{Goodman}},
  \emph{\bibinfo{title}{Introduction to Fourier Optics}}
  (\bibinfo{publisher}{McGraw-Hill}, \bibinfo{year}{1996}).

\bibitem[{\citenamefont{Dunning and Hulet}(1996)}]{dunn96}
\bibinfo{editor}{\bibfnamefont{F.~B.} \bibnamefont{Dunning}} \bibnamefont{and}
  \bibinfo{editor}{\bibfnamefont{R.~G.} \bibnamefont{Hulet}}, eds.,
  \emph{\bibinfo{title}{Atomic, Molecular, and Optical Physics: Atoms and
  Molecules}} (\bibinfo{publisher}{Academic Press}, \bibinfo{year}{1996}).

\bibitem[{\citenamefont{Zhou and Spruch}(1995)}]{ZhS95}
\bibinfo{author}{\bibfnamefont{F.}~\bibnamefont{Zhou}} \bibnamefont{and}
  \bibinfo{author}{\bibfnamefont{L.}~\bibnamefont{Spruch}},
  \bibinfo{journal}{Phys. Rev. A} \textbf{\bibinfo{volume}{52}},
  \bibinfo{pages}{297} (\bibinfo{year}{1995}).

\bibitem[{\citenamefont{Boyd}(1992)}]{boyd92}
\bibinfo{author}{\bibfnamefont{R.~W.} \bibnamefont{Boyd}},
  \emph{\bibinfo{title}{Nonlinear Optics}} (\bibinfo{publisher}{Academic
  Press}, \bibinfo{year}{1992}).

\bibitem[{\citenamefont{Derevianko et~al.}(1999)\citenamefont{Derevianko,
  Johnson, Safranova, and Baab}}]{DJS99}
\bibinfo{author}{\bibfnamefont{A.}~\bibnamefont{Derevianko}},
  \bibinfo{author}{\bibfnamefont{W.}~\bibnamefont{Johnson}},
  \bibinfo{author}{\bibfnamefont{M.}~\bibnamefont{Safranova}},
  \bibnamefont{and} \bibinfo{author}{\bibfnamefont{J.}~\bibnamefont{Baab}},
  \bibinfo{journal}{Phys. Ref. Lett.} \textbf{\bibinfo{volume}{82}},
  \bibinfo{pages}{3589} (\bibinfo{year}{1999}).

\bibitem[{\citenamefont{Marinescu et~al.}(1997)\citenamefont{Marinescu,
  Dalgarno, and Baab}}]{MDB97}
\bibinfo{author}{\bibfnamefont{M.}~\bibnamefont{Marinescu}},
  \bibinfo{author}{\bibfnamefont{A.}~\bibnamefont{Dalgarno}}, \bibnamefont{and}
  \bibinfo{author}{\bibfnamefont{J.}~\bibnamefont{Baab}},
  \bibinfo{journal}{Phys. Rev. A} \textbf{\bibinfo{volume}{55}},
  \bibinfo{pages}{1530} (\bibinfo{year}{1997}).

\bibitem[{\citenamefont{Lifshitz}(1956)}]{Lif56}
\bibinfo{author}{\bibnamefont{Lifshitz}}, \bibinfo{journal}{JETP}
  \textbf{\bibinfo{volume}{73}} (\bibinfo{year}{1956}).

\bibitem[{\citenamefont{Meystre and Sargent}(1998)}]{MeS98}
\bibinfo{author}{\bibfnamefont{P.}~\bibnamefont{Meystre}} \bibnamefont{and}
  \bibinfo{author}{\bibfnamefont{S.}~\bibnamefont{Sargent}},
  \emph{\bibinfo{title}{Elements of quantum optics}} (\bibinfo{year}{1998}).

\bibitem[{\citenamefont{Ekstrom et~al.}(1995)\citenamefont{Ekstrom,
  Schmiedmayer, Chapman, Hammond, and Pritchard}}]{ESC95}
\bibinfo{author}{\bibfnamefont{C.}~\bibnamefont{Ekstrom}},
  \bibinfo{author}{\bibfnamefont{J.}~\bibnamefont{Schmiedmayer}},
  \bibinfo{author}{\bibfnamefont{M.}~\bibnamefont{Chapman}},
  \bibinfo{author}{\bibfnamefont{T.}~\bibnamefont{Hammond}}, \bibnamefont{and}
  \bibinfo{author}{\bibfnamefont{D.~E.} \bibnamefont{Pritchard}},
  \bibinfo{journal}{Phys. Rev. A} \textbf{\bibinfo{volume}{51}},
  \bibinfo{pages}{3883} (\bibinfo{year}{1995}).

\bibitem[{\citenamefont{Grisenti et~al.}(2000)\citenamefont{Grisenti,
  Schollkopf, Toennies, Manson, Savas, and Smith}}]{gris00}
\bibinfo{author}{\bibfnamefont{R.~E.} \bibnamefont{Grisenti}},
  \bibinfo{author}{\bibfnamefont{W.}~\bibnamefont{Schollkopf}},
  \bibinfo{author}{\bibfnamefont{J.~P.} \bibnamefont{Toennies}},
  \bibinfo{author}{\bibfnamefont{J.~R.} \bibnamefont{Manson}},
  \bibinfo{author}{\bibfnamefont{T.~A.} \bibnamefont{Savas}}, \bibnamefont{and}
  \bibinfo{author}{\bibfnamefont{H.~I.} \bibnamefont{Smith}},
  \bibinfo{journal}{Phys. Rev. A} \textbf{\bibinfo{volume}{61}},
  \bibinfo{pages}{033608} (\bibinfo{year}{2000}).

\end{thebibliography}
\end{document}